# Fingerprint Verification based on Gabor Filter Enhancement

Lavanya B N*, K B Raja* and Venugopal K R*
*Department of Computer Science and Engineering
University Visvesvariah College of Engineering, Bangalore University, Bangalore 560 001
Lavanya_hulkoti@yahoo.com

*Abstract*—Human fingerprints are reliable characteristics for personnel identification as it is unique and persistence. A fingerprint pattern consists of ridges, valleys and minutiae. In this paper we propose Fingerprint Verification based on Gabor Filter Enhancement (FVGFE) algorithm for minutiae feature extraction and post processing based on 9-pixel neighborhood. A global feature extraction and fingerprints enhancement are based on Hong enhancement method which is simultaneously able to extract local ridge orientation and ridge frequency. It is observed that the Sensitivity and Specificity values are better compared to the existing algorithms.

I. INTRODUCTION

The term Biometrics relates to the measurement (metric) of characteristics of a living (Bio) thing in order to identify a person. Biometric recognition is used as an automatic recognition of individuals based on the physiological or behavioral characteristics. A physiological characteristic such as Fingerprint, Face, Iris, Hand geometry and Retina remains same throughout the lifetime of a person. Behavioral characteristics such as signature, gait, voice and keystroke changes with age and mentality of a person.

Fingerprint is one of the most widely used biometric because individuals have tried a number of methods to obliterate or remove their fingerprints by abrading or trying to rub off their fingerprints, burning the fingers, trying to dissolve their fingerprints with strong acids, covering their fingertips with superglue, and even having doctors transplant skin from one finger to another unsuccessfully. The fingerprint is formed in third and fourth month of foetal development and unique with epidermal ridges, furrows and patterns. Distinctiveness and persistence are the highly desirable qualities of fingerprint. Factors such as Skin condition, finger pressure and noise detroit the quality of fingerprint images. The presence of noise gives an image a mottled, grainy, textured or snowy appearance. The images of fingerprint are mistaken during termination; hence it is necessary to investigate the images of fingerprint by using suitable filters. Quality of fingerprint is assessed by Character, Fidelity and Utility.

A smoothly flowing pattern formed by alternating crests (ridges) and troughs (valleys) on the palmer aspect of hand is palm print. Formation of a palm print depends on the initial conditions of the embryonic mesoderm from which they develop. The pattern on pulp of each terminal phalanx is considered as an individual pattern and is commonly referred to as a Fingerprint. A fingerprint is believed to be unique to each person, even identical twins have different. It is one of the most mature biometric technologies and is considered legitimate proofs of evidence all over the world, hence used in forensic divisions worldwide for criminal investigations. Fingerprint-based identification has better matching performance than any other existing biometric technology. The watermark is embedded into the region of interest in the Fingerprint image to enhance the security of Biometric System.

An automatic fingerprint matching is popular in systems which control access to physical locations, computer/network resources, bank accounts or register employee attendance time in company. The matching between the fingerprint pattern are highly sensitive to errors such as various noises, damaged fingerprint areas, the finger being placed in different areas of fingerprint scanner window with different orientation angles and finger deformation during the scanning procedure etc.

Minutiae in the fingerprint images are the terminations and bifurcations of ridge lines. To achieve high accuracy in minutiae of fingerprint images various algorithm is used and they are (i) segmentation algorithm which separate fingerprint image from any noisy background without excluding any ridge valley regions, (ii) image enhancement algorithm preserves the original ridge flow pattern from altering, (iii) minutiae detection algorithm which locates the minutiae points. Challenge is made to show that the three levels of information about the parent fingerprint can be elicited from the minutiae template alone, i.e. the orientation field information, the class or type information and the friction ridge structure.

Experiments using commercial fingerprint matcher suggest that the reconstructed ridge structure bears close resemblance to the parent fingerprint. Fingerprint is represented as graph in which minutiae acts as vertex set and binary minutiae relations provide edge set. Transformation-invariant and transformation-variant features are extracted from binary relation. Local matching probability is estimated from transformation-invariant features and fingerprint rotation transformation with the adaptive Parson window is modeled by transformation-variant features.

Digital camera is used to present a touch-less fingerprint recognition system. Fingerprint images that are acquired using digital camera consist of certain constraints such as low contrast between the ridges and the valleys, defocus and motion blurriness. Touch-less fingerprint recognition system comprises of preprocessing, feature extraction and matching stages. Preprocessing comprises of promising results in terms of segmentation, enhancement and core point detection. Feature extraction is done by Gabor filter and matching by Support Vector Machine (SVM).

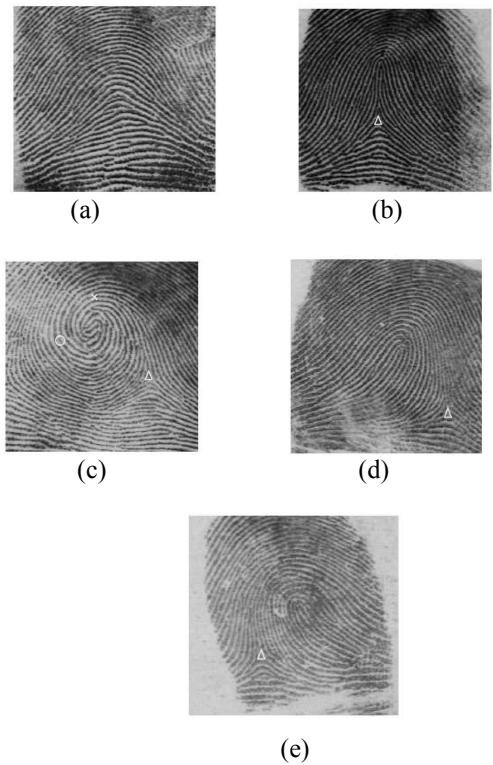

Figure 1. Five classes of Fingerprints (a) Arch (b) Tented Arch (c) Whorl (d) Left Loop (e) Right Loop.

Fingerprint classification is a technique to assign a fingerprint into one of the several pre-specified types already established in the literature which can provide an indexing mechanism. The five major classes of fingerprints are whorl, right loop, left loop, arch, and tented arch as shown in the Figure 1.

*Contribution:* In this paper, the fingerprint image is enhancement by ridge orientation, ridge frequency and Gabor Filter. Minutiae feature extraction is based on 9-pixel neighborhood method.

*Organization:* The paper is organized into the following sections. Section 2 is an overview of related work. The FVGFE model is described in Section 3. Section 4 discusses the algorithm for FVGFE system. Performance analysis of the system is presented in Section 5 and Conclusions are contained in Section 6.

## II. RELATED WORK

Michael Kucken and A. C. Newell [1], discussed the hypothesis on the development of epidermal ridges viz., (i) The epidermal ridge pattern is established as the result of buckling instability acting on the basal layer of the epidermis and resulting in the primary ridges. (ii) The buckling process underlying fingerprint development is controlled by the stresses formed in the basal layer and not by the curvatures of the skin surface. (iii) The stresses that determine ridge direction are themselves determined by boundary forces acting at creases and nail furrow, normal displacements which are most pronounced close to the ridge.

Daesung Moon et. at., [2] implemented the fuzzy fingerprint vault system to secure critical data with fingerprint data. Geometric Hashing and RS decoding algorithms are implemented. Anil Jain et. al., [3] have designed and implemented an on-line fingerprint verification system which operates in two stages: minutiae extraction and minutiae matching. An alignment-based elastic matching algorithm is proposed for minutiae matching. The matching algorithm has the ability to adaptively compensate for the nonlinear deformations and inexact pose transformations between different fingerprints.

Shlomo Greenberg et. al., [4] proposed two methods for fingerprint image enhancement. The first one is carried out using local histogram equalization, Wiener filtering, and image binarization. The second method use a unique anisotropic filter for direct grayscale enhancement. Seifedine Kadry and Aziz Barbar [5] designed Mobile handheld devices which consist of a technique to solve the authenticity problem in mobile communication. It resists any forge imposed by another party.

Tabassam Nawaz et. at., [6] proposed a system that maintains the whole process of taking attendance and maintaining records. The system obtains attendance electronically using sensor and all records are saved on a computer server. Michael Kaas and A. Witkin [7] discussed fingerprint oriented patterns to decompose into two parts: a flow field, describing the direction of anisotropy, and the residual pattern obtained by describing the image in a coordinate system built, from the flow field.

Asker M. Bazen and Gerez [8] presented methods for the estimation of a high resolution directional field from fingerprints. The directional field detects the singular points and the orientations of the points are obtained.

Lin Hong et. al., [9] introduced a fingerprint enhancement algorithm which decomposes the input fingerprint image into a set of filtered images. The orientation field is estimated and a quality mask distinguishes the recoverable and unrecoverable corrupted regions in the input image are generated. Using the estimated orientation field, the input fingerprint image is adaptively enhanced in the recoverable regions.

B. G. Sherlock et. al., [10] described a method of enhancing fingerprint images, based upon non stationary directional Fourier domain filtering. Fingerprints are first smoothed using a directional filter whose orientation is everywhere matched to the local ridge orientation by threshold.

Yun and Cho [11] proposed an adaptive preprocessing method, which extracts five features from the fingerprint images, analyzes image quality with clustering method, and enhances the images according to their characteristics. The preprocessing is performed after distinguishing the fingerprint image quality according to its characteristics.

Brankica M. Popoví´c and L. Ma'skovi c [12] used multiscale directional information obtained from orientation field image to filter the spurious minutiae. The goal of feature extraction in pattern recognition system is to extract information from the input data and depends greatly on the quality of the images. Multiscale directional information estimation is based on orientation field estimation.

F. A. Afsar et. al., [13] presented the minutiae based Automatic Fingerprint Identification Systems. The technique is based on the extraction of minutiae from the thinned, binarized and segmented version of a fingerprint image. The system uses fingerprint classification for indexing during fingerprint matching which greatly enhances the performance of the matching algorithm.

G. Jagadeeswar Reddy et. at., [14] presented fingerprint denoising using both wavelet and curvelet transforms. The search-rearrangement method performs better than minutiae based matching for fingerprint binary constraint graph matching since it does not require implicit alignment of two fingerprint images. K. Zebbiche and F. Khelifi [15] presented biometric images as one Region of interest that has the data processed by most biometric based system. The schemes proposed here consist of embedding the watermark into region of interest in fingerprint images. Discrete Wavelet Transform and Discrete Fourier Transform is used in this algorithm.

Bhupesh Gour et. at., [16] introduced midpoint ridge contour representation inorder to extract the minutiae from fingerprint images. Colour coding scheme is used to scan each ridge only once. K. Karu et. al., [17] proposed fingerprint classification algorithm. Here classification is based on the number and locations of the detected singular points. The algorithm consists of three major steps: (i) computation of the ridge directions, (ii) finding the singularities in the directional image, and (iii) classification of the fingerprint based on the detected singular points.

L.Hong et. at., [19] incorporated a fingerprint enhancement algorithm in the minutiae extraction module. Fast fingerprint enhancement algorithm can improve the clarity of ridge and furrow structures based on estimated local ridge orientation and frequency.

III. MODEL

*A. Definitions*

*1) Fingerprint:* Impression of a finger acquired from digital scanners.

*2) Minutiae:* Ridge bifurcations and ridge endings in fingerprint image.

*3) Core:* Centre of the loop or pattern in fingerprint image. It is located where the innermost recurve begins and curve to exit the same way they came in.

*4) Delta:* It is the area of pattern where there is a triangulation or a dividing of ridges.

*5) Loop:* The fingerprint image contains only one delta.

*6) Whorl***:** The fingerprint image contains 2 or more deltas.

*7) Sensitivity:* The ability of the algorithm to detect the true minutiae and is represented as, 1 − Missed minutiae / ground truth number minutiae.

*8) Specificity*: The ability of the algorithm to reject false minutiae and is represented as, 1 – false minutiae / ground truth number minutiae.

*9) Spurious Minutiae:* It is the type of error that falsely identifies a noisy ridge structure as minutiae.

*10) Missed Minutiae:* It is the type of error that occurs in failing to detect the existing minutiae when it is obscured by surrounding noise, scars or poor ridge structures.

*11) False Minutiae:* Points which have been incorrectly identified as minutiae.

*12) Orientation Image:* The orientation Image represents an intrinsic property of the fingerprint images and defines invariant coordinates for ridges and furrows in a local neighborhood by Least Mean Square orientation estimation algorithm

*B. Block Diagram of FVGFE*

Figure 2 gives the block diagram of FVGFE to extract the Minutiae points for fingerprint verification to disclose personnel identity.

*1) Fingerprint Image:* A gray-level fingerprint image (I) of size N*N is considered to test the FVGFE algorithm. The fingerprint images are scanned at a resolution of 500 dots per inch (dpi).

*2) Fingerprint Image Enhancement:* Fingerprint Enhancement is used to improve the image quality by removing noise by low-pass filter. The configuration of parallel ridges and furrows with well-defined frequency and orientation in a fingerprint image provide useful information which helps in removing undesired noise. The sinusoidal-shaped waves of ridges and furrows vary slowly in a local constant orientation. The band pass filter i.e. tuned to the corresponding frequency and orientation can effectively remove the undesired noise and preserve the true ridge and furrow structures. The region mask is obtained by classifying each block in the fingerprint image into a recoverable or unrecoverable block and If the percentage of recoverable regions is smaller than a threshold, the fingerprint image is rejected. An accepted image is then passed through the filtering stage. Gabor filters have both frequency-selective and orientation-selective properties and have optimal joint resolution in both spatial and frequency domains. Therefore, it is appropriate to use Gabor filters as band pass filters to remove the noise and preserve true ridge/valley structures.

*3) Binarization:* Binarization is the process where the enhanced grayscale image is converted into a binary image. A gray scale image is one which has a specific number of gray levels. For an 8 bit gray scale image can represent $2^8 - 1 = 255$ intensity or gray levels. Most of the fingerprint images are stored as 8 bit gray scale images usually in a bitmap image or as a TIFF image. Due to the large number of gray levels, processing complexity increases. To overcome this, the image is converted into a Binary image.

A binary image is stored as 1-bit image. Usually the grayscale image is converted into a binary image using a threshold value. Let *I(x, y)* represent the intensity value of the enhanced grayscale image at the pixel position *(x, y)*. Let $T_P$ be the threshold value. In case of fingerprint images, $T_P$ represents the differentiating intensity between the background pixels and the ridge pixels. *BW(x, y)* represent the binary image obtained by the Equation (1).

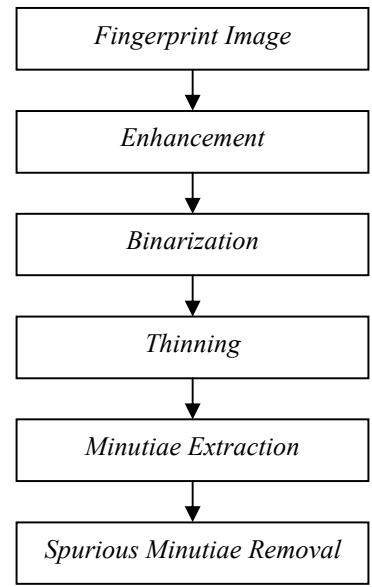

Figure 2. Block Diagram of FVGFE

$$BW(x,y) = \begin{cases} 1 & \text{if } I(x,y) \geq T_p \\ 0, & \text{Otherwise,} \end{cases} \quad (1)$$

*4) Thinning:* Thinning is achieved on the binarized image by morphological operations. A particular morphological operation called dilation is applied to the image until no further changes occur. At this point, the fingerprint image is thinned to a ridge thickness of 1- pixel width. The advantages of thinning: Processing is faster, Complexity is reduced, Memory requirements are reduced and Execution time is less.

Using this transformation, the ridge pixels are converted to white pixel and background is converted to black pixels. The original fingerprint image, Enhanced Image, Binarized Image and Thinned Image are shown in the Figures 3 (a), 3(b),3(c),3(d).

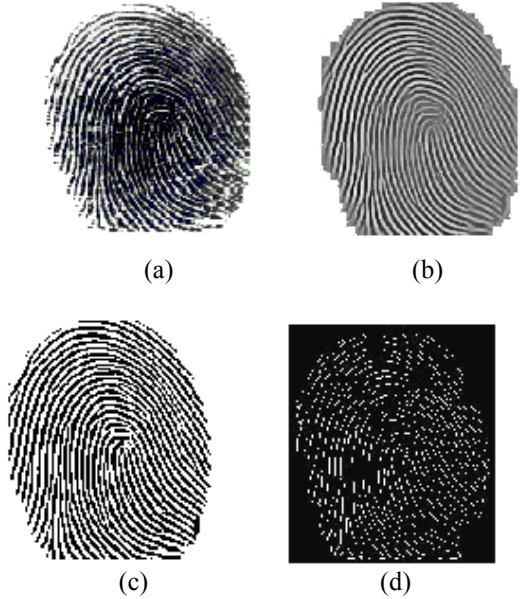

Figure 3. (a) Input Image (b) Enhanced Image (c) Binarized Image (d) Thinned Image.

*5) Minutiae extraction:* Two types of minutiae viz., Ridge bifurcation and Ridge ending are considered. The minutiae are extracted based on the number of pixels in the 9 – pixel neighborhood.

*a) Ridge Bifurcation:* In ridge bifurcation, the number of pixels in the 9 pixel neighborhood is 4 as shown in Figure 4. The X represents a valid minutiae point and 1s in the 9 – pixel window represent the neighboring ridge pixels. Figure 4 (a) and 4 (b) shows an example of vertical ridge bifurcation and horizontal ridge bifurcation respectively.

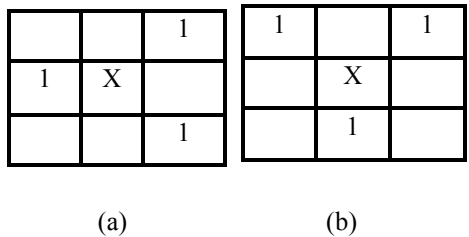

(a)  (b)

Figure 4. Ridge bifurcation

*b) Ridge Ending:* In ridge end, the number of pixels in the 9 pixel neighborhood is 2.

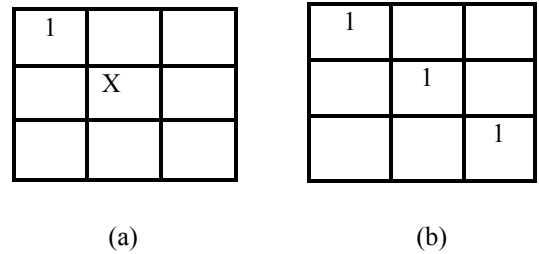

(a)  (b)

Figure 5. Ridge End.

Figure 5 (a) shows a ridge ending with 2 pixels in the 9 pixel neighborhood. Figure 5 (b) shows a normal ridge pixel. All the ridge pixels are characterized by 3 ones in the 9 – pixel neighborhood to differentiate between ridge pixels and minutiae points.

TABLE I.   **NUMBER OF PIXELS IN THE 9- PIXEL NEIGHBORHOOD**

| **Ridges** | Parameters | | |
|---|---|---|---|
| | *Ridge end* | *Ridge bifurcation* | *Ridge Pixel* |
| **Numbers** | 2 | ≥4 | 3 |

The number of pixels in the 9- pixel neighborhood for Ridge End, Ridge Bifurcation and Ridge Pixel are 2, 4 and 3 respectively shown in the Table I.

The redundant minutiae are reduced by (i) All the minutiae points adjacent to each other and within a pre – specified window is removed. (ii) All the minutiae points near the border and within a certain fixed distance from it are removed. (iii) If two ridge endings are encountered close to each other and if no ridges pass through them, and then they are reconnected. Spurious minutiae results from spikes and glitches in the thinned ridge are removed by (i) the distance from the ridge ending to the ridge bifurcation is found. (ii) If the distance is less than or equal to 6, then the minutiae is removed.

IV.   ALGORITHMS

A. *Problem definition*: Given a Fingerprint Image, using FVGFE algorithm, minutiae extraction is carried out for Fingerprint verification.

B. *The Objectives are:*

*a)* The filters are used to improve the quality of fingerprint.

*b)* Binarization and Thinning are performed using threshold and dilation.

  *c)* Minutiae are extracted.

C. *Assumptions:*

  *a)* Fingerprint database of fvc 2004 is considered

  *b)* Scanned fingerprints images are used.

Figure 6 shows the Algorithm of FVGFE for Fingerprint verification**.**

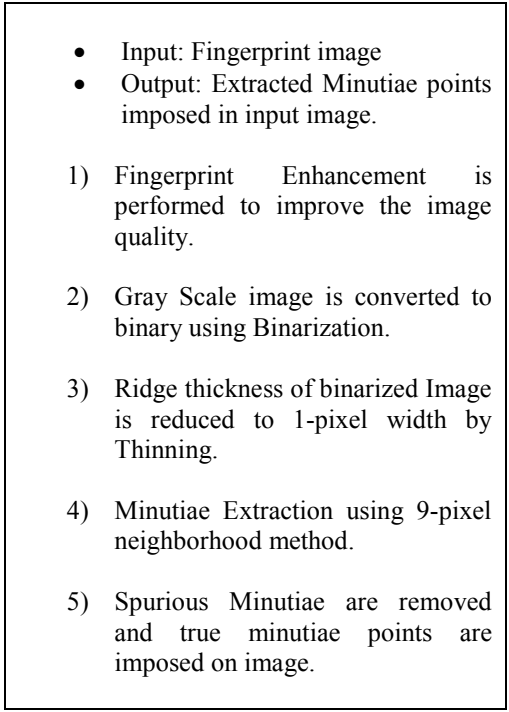

- Input: Fingerprint image
- Output: Extracted Minutiae points imposed in input image.

1) Fingerprint Enhancement is performed to improve the image quality.

2) Gray Scale image is converted to binary using Binarization.

3) Ridge thickness of binarized Image is reduced to 1-pixel width by Thinning.

4) Minutiae Extraction using 9-pixel neighborhood method.

5) Spurious Minutiae are removed and true minutiae points are imposed on image.

Figure 6: Algorithm of FVGFE

## V. PERFORMANCE ANALYSIS

For performance analysis, the Fingerprint database of FVC2004 (DB1) [18] is considered to establish the ground truth of the minutiae. The minutiae features are detected using FVGFE algorithm, the Sensitivity (SEN) and Specificity (SPE) are measured over the database. The SEN and SPE indicate the ability of the algorithm to detect the true minutiae and reject false minutiae respectively.

TABLE II. **MEAN AND STANDARD DEVIATION OF SEN AND SPE**

|  | FEFD | | FVGFE | |
|---|---|---|---|---|
|  | *SEN* | *SPE* | *SEN* | *SPE* |
| **Mean** | 79.8 | 85.36 | 80.8 | 87.29 |
| **S D** | 7.21 | 12.64 | 7.98 | 13.14 |

Table II gives the Mean and Standard Deviation (SD) of sensitivity and specificity for existing Fingerprint Extraction using Fourier Domain Analysis (FEFD) and proposed FVGFE algorithms. It is observed that the values of Sensitivity and Specificity are more in the case of proposed algorithm compared to the existing algorithm

## VI. CONCLUSION

In this paper, we proposed FVGFE algorithm for extraction of minutiae. Fingerprint pattern is enhancement to improve the clarity of the ridge and furrow structures by local ridge orientation and ridge frequency. The minutiae points are extracted from the fingerprint image using 9-pixel neighborhood method. The performance of proposed method is better than the existing method in terms of sensitivity and specificity. In future the minutiae extraction algorithm is combined with ridge extraction for better performance.

## AUTHORS PROFILE

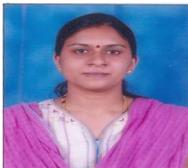

Lavanya B.N. received her Bachelor of Engineering in Computer Science and Engineering from Sri Jayachamarajendra College of Engineering Mysore and Master of Engineering in computer science from Dr.Ambedkar Institute of Technology Banglore.She is persuing her Ph.D in Computer Science and Engineering in JNTU Hyderabad and her area of interest is biometrics, image processing.

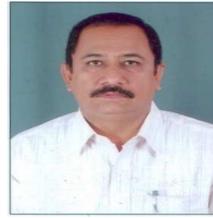

K.B. Raja is an Assistant Professor in Dept of E&CE University Visvesvaraya college of Engg, Bangalore University, Bangalore. He obtained his Bachelor of Engineering and Master of Engineering in Electronics and Communication Engineering from University Visvesvaraya College of Engineering, Bangalore. He was awarded Ph.D. in Computer Science and Engineering from Bangalore University. He has over 35 research publications in refereed International Journals and Conference Proceedings. His research interests include Image Processing, Biometrics, VLSI Signal Processing, computer networks.

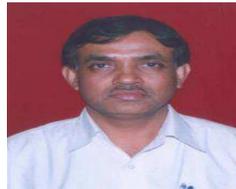

K R Venugopal is currently the Principal and Dean, Faculty of Engineering, University Visvesvaraya College of Engineering, Bangalore University, Bangalore. He obtained his Bachelor of Engineering from University Visvesvaraya College of Engineering. He received his Masters degree in Computer Science and Automation from Indian Institute of Science Bangalore. He was awarded Ph.D. in Economics from Bangalore University and Ph.D. in Computer Science from Indian Institute of Technology, Madras. He has a distinguished academic career and has degrees in Electronics, Economics, Law, Business Finance, Public Relations, Communications, Industrial Relations, Computer Science and Journalism. He has authored 27 books on Computer Science and Economics, which include Petrodollar and the World Economy, C Aptitude, Mastering C, Microprocessor Programming, Mastering C++ etc. He has been serving as the Professor and Chairman, Department of Computer Science and Engineering, University Visvesvaraya College of Engineering, Bangalore University, Bangalore. During his three decades of service at UVCE he has over 200 research papers to his credit. His research interests include computer networks, parallel and distributed systems, digital signal processing and data mining.